\documentclass[%
 reprint,
 amsmath,amssymb,
 aps,
 pra,
floatfix,
]{revtex4-1}

\usepackage{graphicx}
\usepackage{dcolumn}
\usepackage{bm}

\usepackage{multirow}

\begin{document}

\preprint{}

\title{Imaging atomic vacancies in commercially available black phosphorus}

\author{J. V. Riffle$^{1}$, C. Flynn$^{1}$, B. St. Laurent$^{1}$, C. A. Ayotte$^{2}$, C. A. Caputo$^{2}$, S. M. Hollen$^{1}$}
\affiliation{$^{1}$Department of Physics, University of New Hampshire, Durham, NH 03824}
\affiliation{$^{2}$Department of Chemistry, University of New Hampshire, Durham, NH 03824}

\date{\today}

\begin{abstract}
Black phosphorus (BP) is receiving significant attention because of its direct 0.4-1.5 eV layer-dependent band gap and high mobility. Because BP devices rely on exfoliation from bulk crystals, there is a need to understand native impurities and defects in the source material. In particular, samples are typically p-doped, but the source of the doping is not well understood. Here, we use scanning tunneling microscopy and spectroscopy to compare atomic defects of BP samples from two commercial sources. Even though the sources produced crystals with an order of magnitude difference in impurity atoms, we observed a similar defect density and level of p-doping. We attribute these defects to phosphorus vacancies and provide evidence that they are the source of the p-doping. We also compare these native defects to those induced by air exposure and show they are distinct and likely more important for control of electronic structure. These results indicate that impurities in BP play a minor role compared to vacancies, which are prevalent in commercially-available materials, and call for better control of vacancy defects.
\end{abstract}

\maketitle
\section{Introduction}
In the new class of 2D van der Waals materials, black phosphorus (BP) has drawn significant attention\cite{Ling2015,Liu2015,Bhimanapati2015,Akinwande2014,Fiori2014,Das2015} because of its layer-dependent direct band gap\cite{Ye2014,Du2010,Das2014} and high mobility of up to $\sim$1000 cm$^2$/Vs at room temperature.\cite{Koenig2014,Li2014,Xia2014,Long2016} These properties put BP in position to bridge between graphene, which has ultrahigh mobility but no band gap, and transition metal dichalcogenides (TMDCs), which have moderate mobilities and band gaps in the optical range. BP additionally has significant structural and electronic anisotropy,\cite{Xia2014,Ye2014} which makes it attractive for new kinds of electronic and optoelectronic applications.\cite{Ling2015,Xia2014,Liu2015}

A significant factor limiting BP device development is in understanding defects and defect creation in BP. For example, it is well-known that BP oxidizes rapidly, making it highly air-sensitive.\cite{Wood2014, Island2015} Experiments either aim to limit time in air or employ encapsulation techniques.\cite{Wood2014,Avsar2015} In addition to its air sensitivity, BP devices and bulk material have been found to be p-doped, but the source of the doping has not been well understood.\cite{Ye2014, Li2014, Morita1986, Keyes1953, Buscema2014,Liu2014a} While some applications can take advantage of this inherent p-doping to make pn junctions or improve Schottky barriers, understanding the source of the doping is essential for achieving material tunability and defect engineering. Efforts that isolate BP by encapsulation have reduced the level of p-doping,\cite{Avsar2015, Buscema2014,Liu2014a, Doganov2015a} but cannot eliminate it. Recent experimental\cite{Kiraly} and theoretical\cite{UmarFarooq2015, Guo2015, Hu2015, Wang2015a} work attributes the p-doping to single phosphorus vacancies. In addition, recent calculations show that phosphorus vacancies oxidize 5000 times faster than perfect lattice sites.\cite{Kistanov2017} To improve the stability and performance of BP for applications, it is essential to understand the prevalence and behavior of these defects. 

In this report, we compare scanning tunneling microscopy and spectroscopy (STM/S) of BP  from two common commercial sources of BP: 2D Semiconductors and HQ Graphene. In samples from both sources we observed a high density of anisotropic, dumbbell-shaped defects. At energies near the band edges, these defects cause $\sim$50 meV inhomogeneities in the local density of states extending to $\sim$10 nm beyond the defect center, and show evidence of charging through tip-induced band bending rings. The strong electronic signature and charging effects are common characteristics of acceptor-type vacancies.\cite{Lee2011,Marczinowski2007,Marczinowski2008,Loth} These data are consistent with recent reports of single phosphorus vacancies in HQ Graphene BP observed by STM and supported by density functional theory.\cite{Kiraly} Additionally, the observation of a similar population of these defects between BP sources with different types of impurity elements points to vacancies as a common source of p-doping. Finally, we compare these native defects to those created by brief air exposure and show they are clearly distinct. At low levels of air exposure, we observe only slight changes in the band gap and doping using $dI/dV$ spectroscopy.  The data indicate that native impurities, and even low levels of air-induced impurities, play a minor role compared to vacancies, which are prevalent in commercially-available BP. These results call for better control of vacancy defects in BP.

\section{Methods}
Our samples were purchased from 2D Semiconductors and HQ Graphene. 2D Semiconductors produces their BP crystals using a diamond anvil cell, starting with red phosphorus. Their impurity concentration is 5 ppm, as measured by secondary ion mass spectroscopy (SIMS) and consists of Li, K, and Na impurities.\cite{privatecomm} HQ Graphene also starts with red phosphorus, but grows their BP crystals using a chemical vapor transport (CVT) method with tin-iodine as a transport gas. The majority impurity from this method is Ti, likely from the quartz tube the BP crystal is grown in, and is 50 ppm measured by SIMS. A secondary impurity is Sn, from the transport gas, but its concentration was too low to be measured by SIMS.\cite{privatecomm}

The samples were stored in a nitrogen glove box with levels of H$_2$O$<$1 ppm and O$_2<$0.1 ppm. We cleaved the samples in inert conditions using adhesive tape. To study pristine surfaces, we compared three preparation methods: 1) Samples are mounted in the N$_2$ glove box on indium at 160$^{\circ}$C. They are transferred to UHV using an air-free transit chamber (home built) that mounts to our UHV system load lock and then into our STM chamber, where they are cleaved \emph{in situ} using carbon tape on a wobble stick; 2) Samples are prepared as in (1), but cleaved in the N$_2$ glove box prior to air-free transfer to the STM stage; 3) Samples are  mounted with conductive epoxy (bulk crystal is exposed to air $\lesssim$ 3 hours) and then cleaved in UHV.  Approach (1) is the most rigorous air-free treatment while (2) and (3) are both easier experimentally. We found no significant differences in the surface structure or defect density observed with samples prepared by these different methods and do not distinguish between them in the remainder of this report. A fourth preparation method was used to compare the pristine surface to one exposed to air. For this comparison, we moved the UHV-cleaved sample to our chamber load-lock and leaked in air to $\sim$200 mbar. After 30 seconds, we brought the load lock back to high vacuum and transferred the sample onto the STM stage.

We imaged the samples in UHV ($\sim 1 \times 10^{-10}$ mbar) using a closed-cycle PanScan Freedom STM from RHK Technology operated at 9 K. We used cut PtIr tips (10\% Pt, 90\% Ir) and calibrated images against atomic resolution images of HOPG. Bias voltages are reported in reference to the sample. We took images in constant current feedback mode, so the image contrast corresponds to apparent height. We used tunneling point spectroscopy to measure the local density of states (LDOS), which is proportional to the derivative of the tunneling current with varying sample bias, $dI/dV$. For the $dI/dV$ spectroscopy reported here, we used a lock-in technique with a 10 mV, 1.3 kHz ac modulation and feedback off. Spatial maps of $dI/dV$ at constant bias were taken simultaneously with topographic images by alternating feedback on and off during imaging. Image analysis was done using WSxM.\cite{Horcas2007}

\begin{figure}
\centering
\includegraphics[width=1\columnwidth]{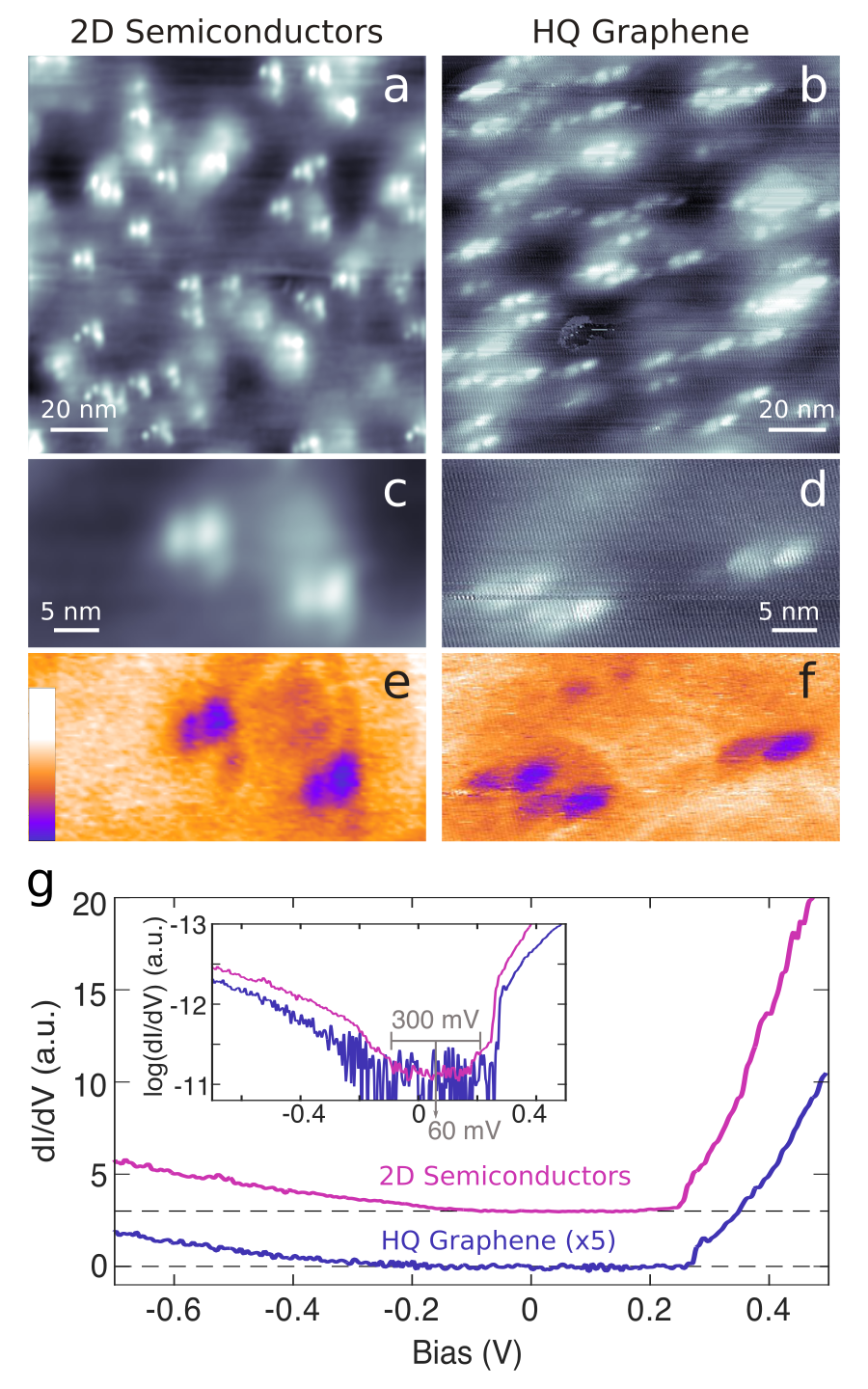} 
\caption{a--b) Large area STM images of BP crystals from 2D Semiconductors and HQ Graphene (-300 mV, 150 pA (a) and -200 mV, 150 pA (b)). c--d) Magnified view of individual defects and e--f) their corresponding dI/dV maps (all at -300 mV, 150 pA). g) dI/dV tunneling point spectroscopy comparison from defect-free regions of each sample (offset for clarity, and HQ Graphene $\times$ 5 to compensate for lower current setpoint ). Bold traces are the average of spectra (shown as lighter traces) at different locations on each sample. Stabilization conditions: $V_{stab}=-1V$;  $I_{stab}= 200  pA$ (2D Semiconductors) and $120 pA$ (HQ Graphene). Inset: same on log-y scale with no baseline offset. }
\end{figure}

\section{STM Comparison of two sources of black phosphorus}
Large area STM images of samples from both sources show a similarly high density of defects (Figure 1a,b).  Defects exhibit a range of apparent heights and spatial extent, consistent with defects appearing on several layers below the surface.\cite{Ebert2003,Jancu2008, Tea2016} Their appearance and density is reproducible for multiple cleaves, supporting that these defects are native and distributed throughout the BP crystal. In filled-state imaging ($V<0$), they exhibit a bright dumbbell shape, most prominent for the top-most crystalline layers, whose long axis is aligned with the armchair ($y$) lattice direction (Figure 1c-f). Recent work by Kiraly \emph{et al.}\cite{Kiraly} shows similar dumbbell defects and attributes them to phosphorus vacancies using DFT simulations. Similar defects were imaged in Qiu \emph{et al.}\cite{Qiu2017}, but attributed to Sn impurities instead. Here, we find these dumbbell defects are present at similar densities in samples from two sources, despite different growth methods that result in different impurity types and an order of magnitude difference in impurity concentrations. The ubiquity of these defects is strong evidence that they are vacancies. 

To compare the electronic differences between these commercial sources of BP, we used tunneling point spectroscopy to probe the local density of states. We find the samples have similar band gaps ($\sim$300 meV) with highly asymmetric band onsets (Figure 1g). A set of such measurements, all taken in regions in-between defects, gives average values of 303$\pm$28 meV and 364$\pm$41 meV for HQ Graphene and 2D Semiconductors BP, respectively (see supplemental material Figure S2,3 for statistics). These values are roughly consistent with theoretical expectations for bulk BP\cite{Rudenko2015} and previous measurements by temperature dependence of Hall conductivity\cite{Keyes1953,Morita1986} and STS.\cite{Zhang2009} The 30-40 meV uncertainty, which is due to large spatial variations in LDOS, makes the $\sim$60 meV difference in band gap between the two samples much less significant. These spatial variations can be attributed to local changes in electronic environment from the vacancy defects, as discussed further below.

We probed carrier doping in $dI/dV$ spectroscopy by measuring the shift of the Fermi level toward the valence band. We consistently observed the band gap center shifted to positive bias, as shown in Figure 1g inset where the shift is $\sim$60 meV. Table I lists the conduction band edges, relative to $V_{bias}=0$, for a set of measurements from each sample. By combining these and the band gap values, we find  doping levels to be  68$\pm$39 meV for 2D Semiconductors and 116$\pm$23 meV for HQ Graphene. This shift indicates inherent p-doping in both samples, as is commonly observed in BP devices,\cite{Keyes1953,Ye2014,Buscema2014a} with a somewhat higher doping in HQ Graphene samples. Again, spatial variations in doping, characterized by the standard deviation of a collection of measurements, are of similar magnitude as the difference in doping between the samples.

\begin{table}
\setlength{\tabcolsep}{0.2em} 
\renewcommand{\arraystretch}{1.2}
\centering
\begin{tabular}{| l  l | c    c |}
 \cline{3-4}
 \multicolumn{2}{c|}{}& HQ Graphene & 2D Semiconductors \\ 
 \hline 
 \multicolumn{2}{| l |}{Zig-zag (x)} & 3.30 $\pm$ 0.06 {\AA}  & 3.28 $\pm$ 0.05 {\AA} \\
 \hline
  \multicolumn{2}{| l |}{Armchair (y)} & 4.48 $\pm$ 0.07 {\AA} & 4.46 $\pm$ 0.07 {\AA} \\
 \hline
  \multicolumn{2}{| l |}{Band gap} & 303 $\pm$ 28 meV & 364 $\pm$ 41 meV \\
 \hline
  \multicolumn{2}{| l | }{Conduction band edge} & 267 $\pm$ 23 meV & 250 $\pm$ 18 meV \\
  \hline
  \multicolumn{2}{| l | }{Effective p-doping} & 116$\pm$23 meV  &  68$\pm$39 meV \\
 \hline
   \multicolumn{2}{| l | }{Defect concentration} &61 $\pm$ 6 ppm & 63 $\pm$ 13 ppm \\
\hline 
\end{tabular}
\caption{Comparison of two commercial sources of BP measured by STM/S. Lattice constants measured from three HQ Graphene images and four 2D Semiconductors images over a range of voltages ($\pm$1.3 V, $\pm$0.5 V, and $\pm$0.3 V). Band gap measurements were sampled over more than 15 different positions on each of the samples. Defect concentrations were calculated for 5 layers (including surface), estimated from apparent height statistics (Figure S5).}
\end{table}

\begin{figure}
\begin{center}
\includegraphics[width=0.9\columnwidth]{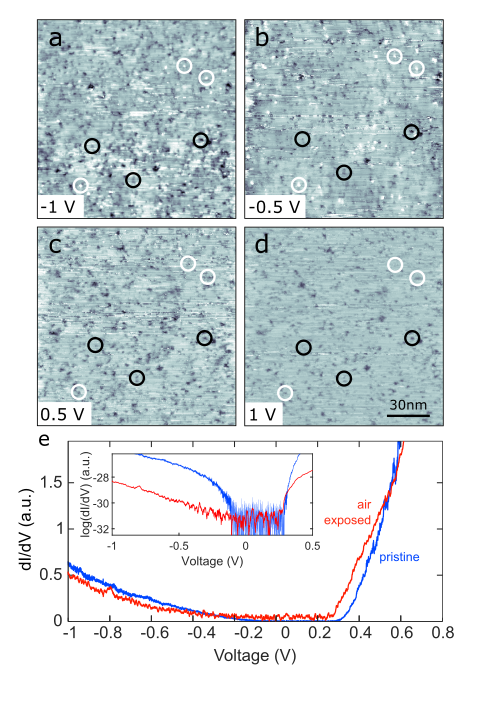}
\caption{a--d) STM topographs of the BP surface at -1V, -0.5V, 0.5V, and 1V after 30 seconds of air exposure at 200 mbar.  Air-induced defects and native vacancy defects are marked as black circles and white circles, respectively. e) Point spectroscopy of BP surface before (blue) and after (red) exposure to air.  Inset: after air exposure on a log scale. The valence band edge is identified as the intersection of the in-gap mean (dashed gray line) with the trend line of valence band (black dashed line). (HQ Graphene sample)}
\end{center}
\end{figure}

Compiling these results reveals that the prevalence of vacancies dominates the electronic behavior of these samples, despite the differences in the material purity and growth methods. Table I summarizes the comparison of lattice constants, band gaps, effective doping, and defect concentrations. Lattice constants are in good agreement with each other and reported values from x-ray powder diffraction\cite{Brown1965} and previous STM results.\cite{Zhang2009, Liang2014} To measure the volumetric density of vacancy defects, we count the defects and estimate the number of crystalline layers represented using apparent height statistics. We examine the distribution of apparent heights within a representative image and find that STM probes a maximum of 5 crystalline layers (Figure S5). This gives a minimum of $\sim$ 60 ppm vacancy defects in both samples. This density is higher than impurity densities measured by SIMS in both samples.\cite{privatecomm} Because of the similar vacancy densities, band gaps, and doping levels between samples with different impurity types and densities, we conclude vacancies are the most likely source of the p-doping observed in BP, consistent with theoretical predictions.\cite{UmarFarooq2015, Hu2015, Wang2015a} This conclusion is also supported by signatures of vacancy charging/discharging, discussed below.

\section{Air-induced defects in black phosphorus}
To compare air-induced defects to native vacancy defects, we intentionally exposed the BP surface to air at $\sim$200 mbar for 30 seconds. Air-induced defects are easily distinguished from native defects.  The air-exposed surface exhibits dark, irregularly shaped clusters at both positive and negative imaging biases, shown in Figure 2 (black circles). This imaging contrast is consistent with the expected insulating nature of the oxidized sites, which results in decreased tunnel current for all biases. Native defects (white circles) are bright at negative biases and dark at positive biases.

Point spectroscopy on defect-free regions of the air-exposed sample shows an increased band gap (from 296 $\pm$ 15 to 392 $\pm$ 36 meV) and decreased p-doping (from 122 $\pm$ 17 to 48 $\pm$ 36 meV) relative to the sample prior to exposure (Figure 2e). These changes are consistent with gated photoluminescence measurements, which show evidence that plasma-induced oxygen defects n-dope BP flakes\cite{Pei2016}. But measurements in BP FETs exposed to air showed further p-doping\cite{Wood2014} and first-principles calculations predict oxygen defects in BP are neutral or p-doping\cite{Ziletti2015,Kistanov2017}. High resolution images and spectroscopy of these air-induced defect complexes could provide further insight into these and related questions regarding the role of vacancies in structural and electronic degradation of BP. 

\section{Defect-induced electronic inhomogeneities in black phosphorus}
The density of the native defects, in combination with their large spatial extent (more than 20 lattice constants) creates a highly inhomogeneous electronic landscape (Figure 3). The defects' apparent height and corresponding contrast in LDOS increases upon approaching the valence band edge from high negative bias (Figure 3a-d and e-h). This spatial inhomogeneity is reflected in the standard deviation of band gaps and doping levels measured at different locations, reported in Table I. 

\begin{figure}
\begin{center}
\includegraphics[width=1\columnwidth]{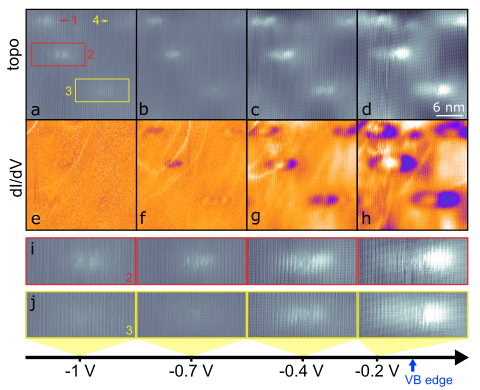}
\caption{ a--d) topographic images and e--h) simultaneously collected $dI/dV$ maps. White (purple) reflects regions with high (low) LDOS. ($V_{bias}, I_{set}$)=$(-1, 200)$, $(-0.7,200)$, $(-0.4,150)$ and $(-0.2V, 100pA)$ from left to right. i--j) Magnified views of defects 2 (red) and 3 (yellow) identified in (a). Vertical lines are zig-zag atomic rows. The valence band edge is indicated at $\approx$0.15 eV. (HQ Graphene sample)}
\end{center}
\end{figure}

\begin{figure}
\centering
\includegraphics[width=.8\columnwidth]{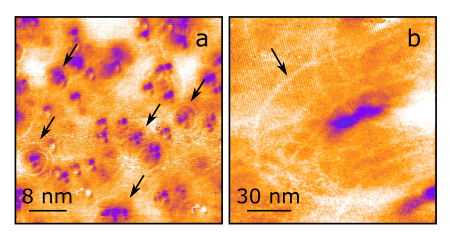}
\caption{$dI/dV$ maps of BP samples showing rings of high LDOS around defects due to tip-induced band bending for a) 2D Semiconductors (-300 mV, 150 pA) and b) HQ Graphene (-350 mV, 150 pA) samples.}
\end{figure}

Figure 3 also shows the dumbbells have a distinctive left/right asymmetry between their lobes, which grows more intense close to the valence band edge. The right lobe of all the defects is brighter than the left lobe, in both the topography and $dI/dV$ maps (Figure 3). Among the four defects imaged in Figure 3, two defects (red labels) are significantly brighter at all biases than the other two (yellow labels), consistent with imaging defects on different crystalline layers. The lobe asymmetry is stronger in surface defects (Figure 3i) but still evident in subsurface defects (Figure 3j). This lobe asymmetry consistently occurs with the same left/right orientation for a population of defects (see \emph{e.g.} Figure 1). The asymmetry in the lobes is expected for vacancies on different lattice sites, $A$ and $B$\cite{Kiraly}, but the uniform population ($i.e.$ the right side is always brighter) indicates a preference for one sublattice site over the other, which requires a broken lattice symmetry. This can be explained by surface buckling, which would make one site more favorable than the other for vacancy formation. This explanation was suggested by Kiraly \emph{et al.}\cite{Kiraly} and supported by theory.\cite{Cai2016,Wang2015a,Guo2015,Ziletti2015} It is also now supported by structural measurements of BP surfaces using LEEM and LEED-IV.\cite{Dai2017} However, it does require that vacancies, which presumably occupy $A$ and $B$ sites evenly in the bulk, move upon cleaving to all \emph{e.g.} $A$ sites. Furthermore, this must happen in the surface layer and 1-2 layers beneath the surface to be consistent with our observations.

In addition to the spatial inhomogeneities, there are signatures in these data that the vacancies are charging and discharging as a result of tip-induced band bending (TIBB)\cite{Pradhan2005,Teichmann2005,Brar2011}. Because the STM tip acts as a local gate, it pulls or pushes the bands in its vicinity, and creates an energy landscape where carriers can tunnel into in-gap defect states at some radius from the defect. This extra impurity tunneling then appears as a ring around defects in $dI/dV$ images. Examples of this signature of defect charging/discharging is shown in Figure 4 for samples from both sources. Because of the high density of defects in these samples, there are many rings overlapping each other, which complicates a more detailed analysis. Nonetheless, these data provide strong support for the assignment of vacancy defects as acceptors and are also consistent with the recent reports of TIBB by Qiu \emph{et al.}\cite{Qiu2017} for HQ Graphene BP.

\section{Conclusion}
These experiments show BP samples from two sources grown by two methods have a similarly high density of phosphorus vacancies, in contrast to their order of magnitude difference in impurity concentration. The samples also show very similar band gaps and doping levels. The vacancies in the samples have a highly inhomogeneous electronic structure and show signatures of charging/discharging at energies near the valence band edge. These results suggest that vacancies are the dominant factor in determining the electronic quality---and specifically the p-doping---in BP. In contrast, defects created by air exposure showed a modest increase in the band gap and decrease in the p-doping. Further investigation of these air-induced defects is needed to explain inconsistencies between theory\cite{Ziletti2015} and experiment\cite{Wood2014} regarding the electronic impact of air-induced defects and the role of vacancies.  Finally, we concur with the evidence\cite{Kiraly,Dai2017,Wood2014,Pei2016} of a sublattice preference for phosphorus vacancies, supporting surface buckling. These results will influence the community working to control the properties of 2D material devices and redirect the 2D crystal growth community, as it is now clear that vacancies are the primary factor affecting the quality of BP source material. 

\section*{Acknowledgements}
We are grateful to the Center for Advanced Materials and Manufacturing Innovations at the University of New Hampshire for their generous seed funding of this work. We are also grateful for insightful discussions with Jay Gupta, Jiadong Zang, and Jie-Xiang Yu.

\bibliography{bibliography}

\end{document}